\documentclass[twoside,10pt]{article}
\usepackage{amsmath}
\usepackage{amssymb}
\usepackage{graphicx}
\usepackage{color}

\def\*{{\phantom *}}
\def\stackunder#1#2{\mathrel{\mathop{#2}\limits_{#1}}}
\def\dsum{\mathop{\displaystyle \sum }}%
\def\dint{\mathop{\displaystyle \int}}%

\newtheorem{theorem}{Theorem}[section]
\newtheorem{lemma}[theorem]{Lemma}

\newtheorem{corollary}[theorem]  {Corollary}
\newtheorem{remark}[theorem]  {Remark}

\renewcommand{\section}{\secdef\sct\sect}
\newcommand{\sct}[2][default]{\refstepcounter{section}
\vspace{0.5cm} \setcounter{equation}{0}
\centerline{ %\large
\scshape \arabic{section}.\ #1} \vspace{0.3cm}}
\newcommand{\sect}[1]{
\vspace{0.5cm} \centerline{\large\scshape #1} \vspace{0.3cm}}

\renewcommand{\subsection}{\secdef \subsct\sbsect}
\newcommand{\subsct}[2][default]{\refstepcounter{subsection}
\nopagebreak \vspace{0.5\baselineskip} {\flushleft\bf
\arabic{section}.\arabic{subsection}~\bf #1  } \nopagebreak}
\newcommand{\sbsect}[1]{\vspace{0.1cm}\noindent
{\bf #1}\vspace{0.1cm}}

\renewcommand{\subsubsection}{%
\secdef \subsubsect\sbsbsect}
\newcommand{\subsubsect}[2][default]{%
\refstepcounter{subsubsection} \nopagebreak
\vspace{0.1\baselineskip} \nopagebreak {\flushleft
\sffamily\slshape
\arabic{section}.\arabic{subsection}.\arabic{subsubsection}
\ %
\sffamily #1\/.}\ }
\newcommand{\sbsbsect}[1]{\vspace{0.1cm}\noindent
{\bf #1}\ }

\begin{document}

\begin{center}

{\Large{\textbf{Large Deviations
in the Superstable Weakly Imperfect Bose Gas}}}

\vspace{0.5 cm}

{J.-B. Bru${}^{a}$ and V.A. Zagrebnov${}^{b}$ }\\
%EndAName
{\normalsize \textit{\hspace{-.5 cm}}}\\
\hspace{-.5 cm}{\normalsize \textit{${}^{a}$Fakult\"at f\"ur Physik,
Universit{\"a}t Wien, Boltzmanngasse 5, 1090 Vienna, Austria}}\\
${}^{b}${\normalsize \textit{Universit\'{e} Aix-Marseille II and Centre de Physique Th\'eorique - UMR 6207 \\
{Luminy-Case 907, F-13288 Marseille, Cedex 09, France}}}

%\date{August 09, 2007}
%\maketitle

\end{center}

\begin{abstract}
The superstable Weakly Imperfect Bose Gas {(WIBG)} was originally derived to solve
the inconsistency of the Bogoliubov theory of superfluidity. Its
grand-canonical thermodynamics was recently solved but not at {point of} the
{(first order)} phase
transition. This paper proposes to close this gap by using the large
deviations formalism and in particular the analysis of the Kac distribution function.
It turns out that, as a function of the chemical
potential, the discontinuity of the Bose condensate density at the phase
transition {point} disappears as a function of the particle density. Indeed, the
Bose condensate continuously starts at the first critical particle density and
progressively grows but the free-energy per particle stays constant until the
second critical density is reached. At higher particle densities, the Bose
condensate density as well as the free-energy per particle both increase {monotonously}.
\end{abstract}

%\vspace{2 cm}

\vspace{0.5 cm}


\vspace{\fill} %\hrule width2truein \smallskip
%{\baselineskip=10pt \noindent Copyright \copyright\ 2001 by the
%authors. Reproduction of this article in its entirety, by any means,
%is permitted for non-commercial purposes.\par }

\renewcommand{\thefootnote}{\fnsymbol{footnote}}

%%%%%%%%%%%%%%%%%%%%%%%%%%%%%%%%%%%%%%%%%%%%%%%%%%%%%%%%%%%%%%%%%%%%%%%%%%%%%%%%%%%%%%%%%%%%%%%%%%
%\tableofcontents
%%%%%%%%%%%%%%%%%%%%%%%%%%%%%%% INTRODUCTION %%%%%%%%%%%%%%%%%%%%%%%%%%%%%%%%%%%%%%%%%%%%%%%%%%%%%

\section{Introduction}

The proof of large deviations for the distribution of the particle
density (the Kac distribution) in the Perfect and in the Mean-Field Bose gases goes
back to \cite{LewisPuleZagrebnov1}. In recent papers \cite
{LebLenSpo,GallLebMast}, the authors addressed to the large deviations in
the particle density in a sub-domain both for the perfect and for rarified quantum
gases (Fermi or Bose). In the present paper we extend the study of {Large Deviations (LD)}
principle to the superstable Weakly Imperfect Bose Gas (WIBG) \cite
{AngelescuVerbeureZagrebnov1}, known also as the Superstable Bogoliubov
model \cite{ZagBru}. The study of this model started in \cite
{AngelescuVerbeureZagrebnov1,AngelescuVerbeureZagrebnov2} was recently
completed in \cite{Adams-Bru1,Adams-Bru2,Bru2006,Adams-Bru1bis}.

Actually, this model originates from a \textit{weaker truncation} than that of the Bogoliubov one
in the grand-canonical ensemble. This new system {served
to solve some inconsistencies between} the grand-canonical Bogoliubov theory of
superfluidity and the WIBG description. This non-diagonal boson model was rigorously solved on the
thermodynamic level for the grand-canonical ensemble in \cite
{Adams-Bru1bis,Bru2006}. It turns out {that similar to the WIBG it manifests} a phase transition
with a \textit{non-conventional} Bose condensation at high densities $\rho $ or high
inverse temperatures $\beta $. Meantime, even for $\beta \uparrow +\infty ,$
i.e. at a zero-temperature, only a fraction of the full density is in the
condensate: i.e. there is a \textit{coexistence} of particles inside and outside the
boson condensate. This last phenomenon is known as  \textit{depletion} of the
condensate. More interesting for our analysis is a discontinuity of the
particle density from $\rho _{-}>0$ to $\rho _{+}>\rho _{-}$ related to a
strictly positive jump of the condensate density at the  phase transition
defined by a fixed chemical potential $\mu _{c}.$ This \textit{first-order} phase
transition as a function of the chemical potential $\mu $ {sounds unusual and
seems to be not quite clear as far as it concerns its physical relevance.}

In fact, the grand-canonical thermodynamics of the superstable WIBG is
\textit{unknown} at {the point of coexistence of the low- and high-density phases.} This
paper proposes to close this gap {using large deviations techniques description of the density distribution.}
For instance {to answer the question, what is the value} of the Bose condensate density
when $\rho \in [\rho _{-},\rho_{+}]$ ?  In fact several scenarios are possible. Since this phase transition
is characterized by the appearance of a {non-conventional Bose condensation, which is due to particle
interaction}, a naive thought might be that there is no condensate at all in domain $\rho \in (\rho
_{-},\rho _{+})$, i.e. the condensate density jumps {from zero} to a strictly positive
value for $\rho >\rho _{+}$. {In fact} this scenario is \textit{wrong}. Here we
show that {this discontinuity is a subtle function} of the total particle density $\rho $. Formally, at
the point of the phase coexistence, the corresponding quantum Gibbs state of the model is no
more a pure state \cite{BrattelliRobinson} but instead, a convex
combination of some of them. {A similar observation was made for example} in Section 4 of \cite
{LewisPuleZagrebnov1}.

Actually, we verify LD for the Bose condensate density for any {given}
particle densities $\rho $ in the grand-canonical ensemble, i.e. even {at the point of the}
phase transition. A direct consequence of this study is a rigorous proof
that the discontinuity of the Bose condensate and its depletion, {visible as a
function of the chemical potential $\mu $, does not appear in the same grand-canonical ensemble
if it is considered as a function of the total particle density $\rho >0$}. We show that
the Bose condensate density continuously increases with $\rho >0$. In others
words, there is no jump and the phase transition in $\rho >0$ is of the second
order. When the particle density $\rho $ (or the inverse temperature $\beta $) exceeds the first critical
value $\rho _{-}$, the Bose condensate density {continuously} grows but the
free-energy per particle, i.e., the corresponding chemical potential $\mu
_{\rho },$ stays constant: $\mu _{\rho }=\mu _{c}$  in domain: $\rho \in
[\rho _{-},\rho _{+}]$. At higher particle densities (or inverse temperatures
$\beta $), the Bose condensate as well as the free-energy per particle $\mu
_{\rho }>\mu _{c}$ both increase when $\rho >\rho _{+}$.

The structure of the paper is {the following.} {In Section \ref{Section WIBG}
we briefly review the grand-canonical thermodynamics of the superstable WIBG for a given particle
density $\rho $}. {Our main results} are
formulated in Section \ref{section results}. The proofs are
collected in Section \ref{proof}. For the reader convenience,
we collect in Appendix (Section \ref{section appendix}) some
technical results as well as a short review on the LD principles.

To conclude, we recall that throughout this paper $\beta >0$ denotes the inverse
temperature, whereas $\mu $ and $\rho >0$ are respectively the chemical
potential and the total particle density. Also, we reserve the notation $\left\langle -\right\rangle
_{H_{\Lambda }}$ for (\textit{finite-volume}) grand-canonical Gibbs state
corresponding to the Hamiltonian $H_{\Lambda }$.

%%%%%%%%%%%%%%%%%%%%%%%%%%%%%%%%%%%%%%%%%%%%%%%%%%%%%%%%%%%%%%%%%%%%%%%%%%%%%%%%%%%%%%%%%%%%%%%%%%

\section{The Superstable Weakly Imperfect Bose Gas\label{Section WIBG}}

%%%%%%%%%%%%%%%%%%%%%%%%%%%%%%%%%%%%%%%%%%%%%%%%%%%%%%%%%%%%%%%%%%%%%%%%%%%%%%%%%%%%%%%%%%%%%%%%%%

\subsection{The Hamiltonian \protect\cite{AngelescuVerbeureZagrebnov1}\label%
{Section Ham}}

\noindent Let an homogeneous gas of $n$ spinless bosons with mass $m$ be
enclosed in a cubic box $\Lambda \subset \Bbb{R}^{3}$ of volume $V:=|\Lambda
|.$ The one-particle energy spectrum is then $\varepsilon _{k}:=\hbar
^{2}k^{2}/2m$ and, using periodic boundary conditions, $\Lambda ^{*}:=(2\pi
\Bbb{Z}/V^{1/3})^{3}\subset \Bbb{R}^{3}$ is the set of wave vectors $k$. The
considered system is with interactions defined via a (real) two-body soft
potential $\varphi (x)=\varphi (||x||)$ such that:

\begin{itemize}
\item[(A)]  $\varphi \left( x\right) \in L^{1}\left( \Bbb{R}^{3}\right) $
(absolute integrability).

\item[(B)]  Its (real) Fourier transformation $\lambda _{k}=\lambda _{||k||}$
satisfies: $\lambda _{0}>0\ $and $0\leq \lambda _{k}\leq \lambda _{0}$ for $%
k\in \Bbb{R}^{3}$.
\end{itemize}

\noindent The Superstable WIBG {(also known as
the AVZ Hamiltonian \cite{Adams-Bru1bis} or the Superstable Bogoliubov Hamiltonian \cite{Adams-Bru1}),
was proposed for the first time in \cite{AngelescuVerbeureZagrebnov1}}. It is defined by
\begin{equation}
H_{\Lambda ,\lambda _{0}>0}^{SB}:=H_{\Lambda ,0}^{B}+U_{\Lambda }^{MF}.
\label{hamiltonian AVZ}
\end{equation}
Here the weakly imperfect Bose gas
\begin{equation}
H_{\Lambda ,0}^{B}:=\stackunder{k\in \Lambda ^{*}\backslash \left\{
0\right\} }{\sum }\left\{ \varepsilon _{k}a_{k}^{*}a_{k}+\frac{\lambda _{k}}{%
2}\left( \frac{a_{0}^{*}a_{0}}{V}\left(
a_{k}^{*}a_{k}+a_{-k}^{*}a_{-k}\right) +a_{k}^{*}a_{-k}^{*}\frac{a_{0}^{2}}{V%
}+\frac{a_{0}^{*2}}{V}a_{k}a_{-k}\right) \right\}
\end{equation}
contains the kinetic-energy term\footnote{%
Recall that $\varepsilon _{0}=0.$} plus diagonal and non-diagonal
interactions. It is solved in the canonical ensemble in \cite
{Adams-Bru2,Bru2006}. The repulsive interaction ensuring the superstability
of $H_{\Lambda ,\lambda _{0}}^{SB}$ by assumptions (A)-(B) is the ``forward
scattering'' interaction
\begin{equation}
U_{\Lambda }^{MF}:=\dfrac{\lambda _{0}}{2V}\stackunder{k_{1},k_{2}\in
\Lambda ^{*}}{\dsum }a_{k_{1}}^{*}a_{k_{2}}^{*}a_{k_{2}}a_{k_{1}}=\dfrac{%
\lambda _{0}}{2V}\left( N_{\Lambda }^{2}-N_{\Lambda }\right) ,\text{ with }%
N_{\Lambda }:=\stackunder{k\in \Lambda ^{*}}{\sum }a_{k}^{*}a_{k}
\end{equation}
defined as the particle-number operator within the grand-canonical
framework. Indeed, $a_{k}^{*}$ and $a_{k}$ are the usual boson creation /
annihilation operators in the one-particle state\footnote{%
Here $\chi _{\Lambda }\left( x\right) $ is the characteristic function of
the box $\Lambda .$} ${\chi _{\Lambda}(x)}
e^{ikx}/{\sqrt{V}}$, acting on the boson Fock space
\begin{equation}
\mathcal{F}_{\Lambda }^{B}:=\stackunder{n=0}{\stackrel{+\infty }{\bigoplus }}%
\mathcal{H}_{B}^{\left( n\right) },\text{ with }\mathcal{H}_{B}^{\left(
n\right) }:=\left( L^{2}\left( \Lambda ^{n}\right) \right) _{\text{symm}},%
\text{ }\mathcal{H}_{B}^{\left( 0\right) }:=\Bbb{C},
\label{boson Fock space}
\end{equation}
defined as the symmetrized $n$-particle Hilbert spaces, see \cite
{BrattelliRobinson,Ruelle}.

\begin{remark}
\label{remark Fock space}Let $\mathcal{H}_{0\Lambda }\subset L^{2}\left(
\Lambda \right) $ be the one-dimensional subspace generated by $\psi
_{k=0}\left( x\right) =1/\sqrt{V}$. Then $\mathcal{F}_{\Lambda }^{B}\approx
\mathcal{F}_{0\Lambda }\otimes \mathcal{F}_{\Lambda }^{\prime }$ where $%
\mathcal{F}_{0\Lambda }$ and $\mathcal{F}_{\Lambda }^{\prime }$ are the
boson Fock spaces constructed out of $\mathcal{H}_{0\Lambda }$ and of its
orthogonal complement $\mathcal{H}_{0\Lambda }^{\bot }$ respectively.
\end{remark}

\subsection{Grand-canonical thermodynamics for a fixed particle density
\protect\cite{Adams-Bru1bis,Adams-Bru2,Bru2006}\label{Section AVZ}}

{We consider here the grand-canonical
ensemble $(\beta, \mu)$ defined by a given particle density $\rho $, or more precisely by the chemical
potential ${\mu}_{\Lambda, \rho}$}, which is a unique solution of the equation
(\ref{mu fixed particle density}) below. In any finite volume,
the corresponding particle density is
strictly increasing by strict convexity of the pressure. Therefore, for any $%
\rho >0,$ there exists a unique $\mu _{\Lambda ,\rho }$ such that
\begin{equation}
\left\langle \frac{N_{\Lambda }}{V}\right\rangle _{H_{\Lambda ,\lambda
_{0}}^{SB}}=\rho ,  \label{mu fixed particle density}
\end{equation}
where $\left\langle -\right\rangle _{H_{\Lambda ,\lambda _{0}}^{SB}}$ always
represents the (finite volume) grand-canonical Gibbs states for $H_{\Lambda
,\lambda _{0}}^{SB}$ taken at inverse temperature $\beta $ and chemical
potential $\mu _{\Lambda ,\rho }.$ In the thermodynamic limit, $\mu
_{\Lambda ,\rho }$ converges to $\mu _{\rho }\in \Bbb{R}$ for any $\rho >0$.
In fact, $\mu _{\rho }$ is strictly increasing except for $\rho \in [\rho
_{-},\rho _{+}]$ where it equals $\mu _{c}=\mu _{c}$. Here $\rho _{+}>\rho
_{-}>0$ are two well-defined density only depending on the inverse
temperature $\beta >0.$ Additionally, $\mu _{\rho }:=\alpha _{\rho }+\lambda
_{0}\rho $ with $\alpha _{\rho }<0$ and $\partial _{\lambda _{0}}\alpha
_{\rho }=0$ for $\rho \in [\rho _{-},\rho _{+}].$

\begin{figure}[h]
\centerline{\includegraphics[angle=0,scale=1]{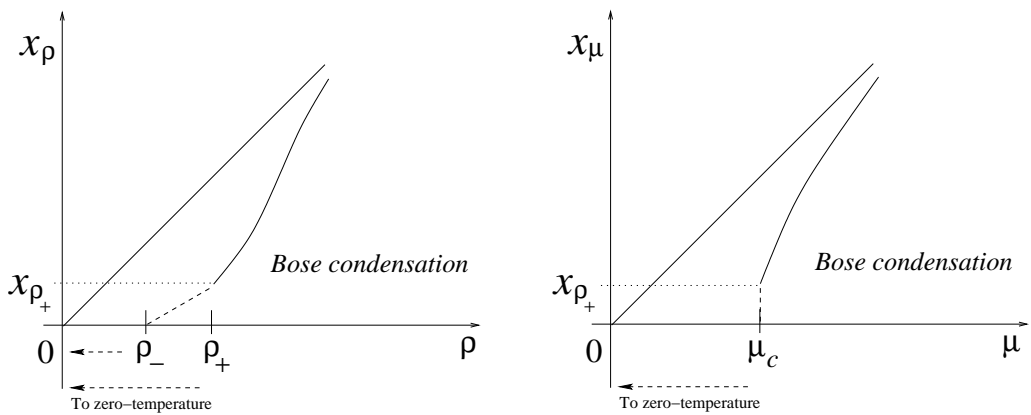}}
\caption{\emph{Illustration of the Bose condensate density, $x_{\rho }$ at fixed particle density $\rho>0$ or
$x_{\mu }$ at fixed chemical potential $\mu \in \Bbb{R}$. The dashed line closing continuously the gap between
$\rho_{-}$ and $\rho_{+}$ in the illustration of $x_{\rho }$ is a consequence of results of the present paper.
{Here each of the asymptotic straight lines are {\rm{:}} $x_\rho = \rho$, or $x_\mu = \mu/\lambda_{0}$. They
correspond to the limits {\rm{:}} $x_{\rho\rightarrow\infty}$, or $x_{\mu\rightarrow\infty}$, with 100\%
of the Bose condensate.}}}
%i.e. from left to right, respectively they are $x=\rho $
%and $\mu =\lambda _{0}x$.}}}
\label{figure-BEC}
\end{figure}

Moreover, there is a non-conventional Bose condensation induced by the
non-diagonal interaction $U_{\Lambda }^{ND}$ for high particle densities:
\begin{equation}
x_{\rho }:=\stackunder{\Lambda }{\lim }\left\langle \dfrac{a_{0}^{*}a_{0}}{V}%
\right\rangle _{H_{\Lambda ,\lambda _{0}}^{SB}}=\left\{
\begin{array}{l}
=0\text{ for }\rho <\rho _{-}, \\
>0\text{ for }\rho >\rho _{+},
\end{array}
\right.  \label{Bose condensate density}
\end{equation}
with $\partial _{\lambda _{0}}x_{\rho }=0$ for $\rho \notin [\rho _{-},\rho
_{+}].$ When $\rho \downarrow \rho _{+}$, note that the Bose condensate
density $x_{\rho }$ converges to $x_{\rho _{+}}>0.$ In particular, since $%
\mu _{\rho }=\mu _{c}$ for $\rho \in [\rho _{-},\rho _{+}],$ the Bose
condensate density $x_{\mu }$ as a function of the chemical potential $\mu $
jumps from $0$ to $x_{\rho _{+}}$ at $\mu =\mu _{c}$. An illustration of the
behavior of $x_{\rho }$ for a fixed density $\rho $ (or $x_{\mu }$ at a
fixed chemical potential $\mu $) is performed in Figure \ref{figure-BEC}.%
%TCIMACRO{
%\TeXButton{newtheory-condensate-melange.eps}{\begin{figure}[h]
% \centerline{\includegraphics[angle=0,scale=1]{newtheory-condensate-melange.eps}}
%\caption{\emph{Illustration of the Bose condensate density, $x_{\rho }$ at fixed particle density
%$\rho>0$ or $x_{\mu }$ at fixed chemical potential $\mu \in \Bbb{R}$. The dashed line closing continuously the
%gap between $\rho_{-}$ and $\rho_{+}$ in the illustration of $x_{\rho }$ will be a consequence of results of this paper.
%Each asymptote corresponds to 100\% of Bose condensate, i.e. from left to right, respectively they are $x=\rho $ and
%$\mu =\lambda _{0}x$.}}
%\label{figure-BEC}
%\end{figure}}}
%BeginExpansion
%EndExpansion

We would like to stress that coexistence of different types of
condensations is a subtle matter. For example a slight modification of
interaction, see \cite{Adams-Bru1bis,Adams-Bru2,Bru2006}, excludes
any coexistence of non-conventional and conventional Bose condensation, as
it appears for high densities in the Bogoliubov WIBG \cite
{BruZagrebnov6,BruZagrebnov8}.

Below we consider coexistence a high and low density phases. For intermediate total density
$\rho \notin [\rho _{-},\rho_{+}]$ we find for the particle density outside the zero-mode
\begin{equation}
\stackunder{\Lambda }{\lim }\dfrac{1}{V}\stackunder{k\in \Lambda
^{*}\backslash \left\{ 0\right\} }{\sum }\left\langle
a_{k}^{*}a_{k}\right\rangle _{H_{\Lambda ,\lambda _{0}}^{SB}}=\frac{1}{%
\left( 2\pi \right) ^{3}}\stackunder{\Bbb{R}^{3}}{\int }\left\{ \frac{f_{k}}{%
E_{k}\left[ e^{\beta E_{k}}-1\right] }+\frac{x_{\rho }^{2}\lambda _{k}^{2}}{%
2E_{k}\left[ f_{k}+E_{k}\right] }\right\} \mathrm{d}^{3}k,
\end{equation}
where $f_{k}:=\varepsilon _{k}-\alpha _{\rho }+x_{\rho }\lambda _{k}$ and $%
E_{k}:=(f_{k}^{2}-x_{\rho }^{2}\lambda _{k}^{2})^{\frac{1}{2}}.$

Observe that the grand-canonical thermodynamic behavior of the superstable
WIBG is unknown for $\rho \in [\rho _{-},\rho _{+}]$ at fixed $\beta >0$.
When $\beta \rightarrow +\infty ,$ i.e. at zero temperature, the critical
densities $\rho _{-}$ and $\rho _{+}$ could both converge to zero depending
on the interaction potential \cite{Adams-Bru1bis,Bru2006}, whereas the
critical chemical potential $\mu _{c}$ converges to a negative value.\
Moreover, we have a non-zero particle density outside the zero-mode for any
fixed $\rho >0$ even at zero-temperature since
\begin{equation}
\stackunder{\beta \rightarrow +\infty }{\lim }\stackunder{\Lambda }{\lim }%
\dfrac{1}{V}\stackunder{k\in \Lambda ^{*}\backslash \left\{ 0\right\} }{%
\dsum }\left\langle a_{k}^{*}a_{k}\right\rangle _{H_{\Lambda ,\lambda
_{0}}^{SB}}>0\mathrm{\ and\ }\stackunder{\beta \rightarrow +\infty }{\lim }%
\stackunder{\Lambda }{\lim }\left\langle \dfrac{a_{0}^{*}a_{0}}{V}%
\right\rangle _{H_{\Lambda ,\lambda _{0}}^{SB}}<\rho .
\end{equation}
In other words, there is a depletion of the Bose condensate even at zero
temperature.

To conclude, the grand-canonical pressure associated with $H_{\Lambda
,\lambda _{0}}^{SB}$ in the thermodynamic limit equals
\begin{eqnarray}
p^{SB}(\beta ,\mu _{\rho }) &=&\stackunder{x\geq 0}{\sup }\left\{
\stackunder{\alpha \leq 0}{\inf }\left\{ p_{0}^{B}(\beta ,\alpha ,x)+\frac{%
(\mu _{\rho }-\alpha )^{2}}{2\lambda _{0}}\right\} \right\}
\label{pressure fixed density_1} \\
&=&\stackunder{\alpha \leq 0}{\inf }\left\{ p_{0}^{B}(\beta ,\alpha ,x_{\rho
})+\frac{(\mu _{\rho }-\alpha )^{2}}{2\lambda _{0}}\right\}
\label{pressure fixed density_2} \\
&=&p_{0}^{B}(\beta ,\alpha _{\rho },x_{\rho })+\frac{\lambda _{0}}{2}\rho
^{2},  \label{pressure fixed density_3}
\end{eqnarray}
for any $\rho >0.$ In particular, $x_{\rho }$ and $\alpha _{\rho }$ are
solutions of the variational problems (\ref{pressure fixed density_1}) and (%
\ref{pressure fixed density_2}) respectively. For any $x\geq 0$ and $\alpha
\leq 0$,
\begin{equation}
p_{0}^{B}\left( \beta ,\alpha ,x\right) :=\stackunder{\Lambda }{\lim }\frac{1%
}{\beta V}\ln \mathrm{Tr}_{\mathcal{F}_{\Lambda }^{\prime }}\left\{
e^{-\beta \left( H_{\Lambda ,0}^{B}\left( x,\alpha \right) -\alpha x\right)
}\right\}  \label{eq 4.26}
\end{equation}
is here the (infinite volume) pressure of the so-called Bogoliubov
approximation\footnote{%
Combined with a gauge transformation $a_{k}\rightarrow e^{i\varphi }a_{k}$}
\begin{equation}
H_{\Lambda ,0}^{B}\left( x,\alpha \right) :=\stackunder{k\in \Lambda
^{*}\backslash \left\{ 0\right\} }{\sum }\left\{ \left( \varepsilon
_{k}-\alpha \right) a_{k}^{*}a_{k}+\frac{x\lambda _{k}}{2}\left(
a_{k}^{*}a_{k}+a_{-k}^{*}a_{-k}+a_{k}^{*}a_{-k}^{*}+a_{k}a_{-k}\right)
\right\}
\end{equation}
of $\{H_{\Lambda ,0}^{B}-\alpha (N_{\Lambda }-a_{0}^{*}a_{0})\}$. Observe
that $H_{\Lambda ,0}^{B}\left( x,\alpha \right) $ is defined on the boson
Fock space $\mathcal{F}_{\Lambda }^{\prime }$ for non-zero momentum bosons,
cf.\ Remark \ref{remark Fock space} and Section \ref{section results} for a
rigorous definition of the so-called Bogoliubov approximation. Finally, note
that the Hamiltonian $H_{\Lambda ,0}^{B}\left( x,\alpha \right) $
represents, via a unitary transformation, a perfect Bose gas of
quasi-particles with one-particle spectrum $E_{k}$ for $k\in \Lambda
^{*}\backslash \left\{ 0\right\} $, see for instance \cite{Adams-Bru1bis}.

%%%%%%%%%%%%%%%%%%%%%%%%%%%%%%%%%%%%%%%%%%%%%%%%%%%%%%%%%%%%%%%%%%%%%%%%%%%%%%%%%%%%%%%%%%%%%%%%%%

\section{Large deviations for the {Bose condensate at a fixed total particle}
density\label{section results}}

To define the (finite volume) distribution $\Bbb{D}_{\Lambda ,\rho }$ of the
condensate, we first recall the rigorous definition of the Bogoliubov
approximation due to Ginibre \cite{Ginibre} and based on coherent vectors.
For any complex $c\in \Bbb{C}$, a coherent vector $|c\rangle $ is an element
of the boson Fock space $\mathcal{F}_{0\Lambda }$ for zero momentum bosons
(cf. Remark \ref{remark Fock space}), satisfying $a_{0}|c\rangle =c\sqrt{V}%
|c\rangle $. In fact, if $\Omega _{0}$ is the vacuum of $\mathcal{F}%
_{\Lambda }^{B},$ then $|c\rangle :=\exp \{-V|c|^{2}/2+c\sqrt{V}%
a_{0}^{*}\}\Omega _{0}$ for any $c\in \Bbb{C}$. The Bogoliubov approximation
of a self-adjoint operator $\mathrm{A}$ acting on $\mathcal{F}_{\Lambda
}^{B} $ is the operator $\mathrm{A}(c)$ defined on the boson Fock space $%
\mathcal{F}_{\Lambda }^{\prime }$ without the zero mode by its quadratic
form
\begin{equation}
\left\langle \psi _{1}^{\prime }\right| \mathrm{A}\left( c\right) \left|
\psi _{2}^{\prime }\right\rangle :=\left\langle c\otimes \psi _{1}^{\prime
}\right| \mathrm{A}\left| c\otimes \psi _{2}^{\prime }\right\rangle ,
\label{definition of Bogo approx}
\end{equation}
for $| c\otimes \psi _{1,2}^{\prime }\rangle$ in the form-domain of $\mathrm{%
A}$.

Now, for any chemical potential $\mu \in \Bbb{R}$ the (finite volume)
grand-canonical pressure associated with $H_{\Lambda ,\lambda _{0}}^{SB}$
equals
\begin{equation}
p_{\Lambda }^{SB}(\beta ,\mu ):=\frac{1}{\beta V}\ln \mathrm{Tr}_{\mathcal{F}%
_{\Lambda }^{B}}\left\{ W_{\Lambda }\right\} ,\mathrm{\ with\;}W_{\Lambda
}:=e^{-\beta (H_{\Lambda ,\lambda _{0}}^{SB}-\mu N_{\Lambda })}.
\label{definition of the pressure}
\end{equation}
By using the generating family of coherent vectors $|c\rangle $ for $c\in
\Bbb{C}$, we can rewrite the trace $\mathrm{Tr}$ above to observe that
\begin{equation}
p_{\Lambda }^{SB}(\beta ,\mu )=\frac{1}{\beta V}\ln \frac{1}{2\pi }%
\stackunder{\Bbb{C}}{\int }\mathrm{Tr}_{\mathcal{F}_{\Lambda }^{\prime
}}\left\{ W_{\Lambda }\left( c\right) \right\} \mathrm{d}^{2}c=\frac{1}{%
\beta V}\ln \frac{1}{2\pi }\stackunder{\Bbb{C}}{\int }e^{\beta Vp_{\Lambda
}^{SB}(\beta ,\mu ,c)}\mathrm{d}^{2}c,
\end{equation}
where $\mathrm{d}^{2}c:=V\pi ^{-1}\mathrm{d}c_{1}\mathrm{d}c_{2}$ with $%
c:=c_{1}+ic_{2}$, $W_{\Lambda }(c)$ results from the Bogoliubov
approximation (\ref{definition of Bogo approx}) of the statistical operator $%
W_{\Lambda },$ and
\begin{equation}
p_{\Lambda }^{SB}(\beta ,\mu ,c):=\frac{1}{\beta V}\ln \mathrm{Tr}_{\mathcal{%
F}_{\Lambda }^{\prime }}\left\{ W_{\Lambda }\left( c\right) \right\}
\label{definition of the pressure as function of c}
\end{equation}
is the pressure defined by the partial trace. For any $\rho >0$, the
corresponding distribution $\Bbb{D}_{\Lambda ,\mu }$ related to the Bose
condensate number density, is now defined on the Borel subsets $\mathcal{A}$
of $\Bbb{C}$ by
\begin{equation}
\Bbb{D}_{\Lambda ,\mu }\left[ \mathcal{A}\right] :=e^{-\beta Vp_{\Lambda
}^{SB}(\beta ,\mu )}\frac{1}{2\pi }\stackunder{\mathcal{A}}{\int }e^{\beta
Vp_{\Lambda }^{SB}(\beta ,\mu ,c)}\mathrm{d}^{2}c.
\label{Kac distribution of the condensate}
\end{equation}
Then, at fixed particle density $\rho >0$, we express a large deviations
principle (Section \ref{Section LDP}) for the condensate distribution $\Bbb{D%
}_{\Lambda ,\rho }:=\Bbb{D}_{\Lambda ,\mu _{\Lambda ,\rho }}$.

\begin{theorem}[{LD principle for the condensate distribution at a fixed density $\rho $}]
\label{theorem LDP-0bis}\mbox{} \newline
The sequence $\{\Bbb{D}_{\Lambda ,\rho }\}$ satisfies a large deviation
principle with speed\textit{\ }$\beta V$ and rate function
\[
\mathrm{D}_{\rho }\left( x\right) :=\stackunder{\alpha \leq 0}{\inf }\left\{
p_{0}^{B}(\beta ,\alpha ,x_{\rho })+\frac{(\mu _{\rho }-\alpha )^{2}}{%
2\lambda _{0}}\right\} -\stackunder{\alpha \leq 0}{\inf }\left\{
p_{0}^{B}(\beta ,\alpha ,x)+\frac{(\mu _{\rho }-\alpha )^{2}}{2\lambda _{0}}%
\right\} ,
\]
for $x=|c|^{2}\geq 0$, cf.\ (\ref{pressure fixed density_1})-(\ref{pressure
fixed density_2}).
\end{theorem}

This theorem shows in particular that the probability to observe a density $%
n_{0}/V\in \mathcal{A}$ of condensed bosons enclosed in $\Lambda $ for a
fixed chemical potential $\mu \in \Bbb{R}$ decreases exponentially with the
volume $V=|\Lambda |$ if the distance between the Bose condensate density $%
x_{\rho }$ (\ref{Bose condensate density}) and the set $\mathcal{A}\subset
\Bbb{R}$ is strictly positive. Now, the next step is to evaluate the
limiting probability measure, in particular at the phase transition defined
for a chemical potential $\rho \in [\rho _{-},\rho _{+}]$. Recall that the
Bose condensate density $x_{\rho }$ (\ref{Bose condensate density})
converges to $0$ when $\rho \uparrow \rho _{-}$ but to a strictly positive
value $x_{\rho _{+}}>0$ when $\rho \downarrow \rho _{+}.$

\begin{theorem}[The condensate distribution outside the point of the phase transition]
\label{theorem LDP-1bis}\mbox{} \newline
The (finite volume) distribution $\Bbb{D}_{\Lambda ,\rho }$ of the
condensate converges weakly in the set of probability measures $\mathcal{M}%
_{1}\left( \Bbb{C}\right) $ as $\Lambda \uparrow \Bbb{R}^{3}$ towards the
singular measures
\[
\Bbb{D}_{\rho }:=\stackunder{\Lambda }{\lim }\Bbb{D}_{\Lambda ,\rho }=\frac{1%
}{2\pi }\stackunder{0}{\stackrel{2\pi }{\int }}\delta \left( c-x_{\rho
}^{1/2}e^{i\theta }\right) \mathrm{d}\theta ,\mathrm{\ }
\]
for any $\rho \in (0,\rho _{-})\cup (\rho _{+},+\infty ).$
\end{theorem}

For $\beta \rightarrow +\infty ,$ i.e. at zero-temperature, observe that $%
\rho _{-}$ and $\rho _{+}$ could both converge to zero, depending on the
interaction potential. But, at finite temperature, i.e. at $\beta >0,$ one
always has $\rho _{+}>\rho _{-}$ and the convergence of $\Bbb{D}_{\Lambda
,\rho }$ is not solved for $\rho \in [\rho _{-},\rho _{+}].$ The
corresponding result is therefore expressed in the next theorem.

\begin{theorem}[The condensate distribution at {the point of} the phase transition]
\label{theorem LDP-1bisbis}\mbox{} \newline
Let $\rho _{+}>\rho _{-}$. As $\Lambda \uparrow \Bbb{R}^{3},$ the (finite
volume) distribution $\Bbb{D}_{\Lambda ,\rho }$ of the condensate converges
weakly in $\mathcal{M}_{1}\left( \Bbb{C}\right) $ towards a convex
combination of the singular measures
\[
\Bbb{D}_{\rho }:=\stackunder{\Lambda }{\lim }\Bbb{D}_{\Lambda ,\rho }=\left(
1-\kappa _{\rho }\right) \delta \left( c\right) +\frac{\kappa _{\rho }}{2\pi
}\stackunder{0}{\stackrel{2\pi }{\int }}\delta \left( c-x_{\rho
_{+}}^{1/2}e^{i\theta }\right) \mathrm{d}\theta ,
\]
for any $\rho \in [\rho _{-},\rho _{+}]$ and with $\kappa _{\rho }:=(\rho
-\rho _{-})/(\rho _{+}-\rho _{-})$.
\end{theorem}

Note that $\kappa _{\rho }$ is a strictly increasing and continuous
function from $[\rho _{-},\rho _{+}]\ $to $\left[ 0,1\right] .$ This result
gives a strong evidence that, at the phase transition, the corresponding Gibbs
state is not a pure state anymore \cite{BrattelliRobinson} but a convex
combination of pure states, see for example Section 4 in \cite
{LewisPuleZagrebnov1}.

Integrating $\Bbb{D}_{\Lambda ,\rho }$ with the function $\varphi
(c)=|c|^{2}$, we finally obtain the Bose condensate density (\ref{Bose
condensate density}) inside the phase transition, i.e. for $\rho \in [\rho
_{-},\rho _{+}]$.

\begin{corollary}[Derivation of the Bose condensate density for any total density]
\label{theorem LDP-1bisbisbis}\mbox{} \newline
The Bose condensate density equals
\[
\stackunder{\Lambda }{\lim }\left\langle \dfrac{a_{0}^{*}a_{0}}{V}%
\right\rangle _{H_{\Lambda _{j},\lambda _{0}}^{SB}}=\left\{
\begin{array}{l}
0\mathrm{\ \qquad \qquad \ \ \ \ \ for\ }\rho \leq \rho _{-}. \\
\dfrac{\rho -\rho _{-}}{\rho _{+}-\rho _{-}}x_{\rho _{+}}\mathrm{\ \ \ for\ }%
\rho \in [\rho _{-},\rho _{+}]. \\
x_{\rho }>0\mathrm{\ \ \ \ \ \ \ \ \ \ \ for\ }\rho \geq \rho _{+}.
\end{array}
\right.
\]
In particular, it is continuous as a function of $\rho >0$ and linearly
increasing for $\rho \in [\rho _{-},\rho _{+}],$ cf. figure \ref{figure-BEC}.
\end{corollary}

As a function of the density $\rho >0$ in the grand-canonical ensemble, the
phase transition is of order two if $\rho _{+}>\rho _{-}$ whereas it is of
order one as a function of the chemical potential. In particular, take $\rho
<\rho _{-},$ then the system behaves as the so-called Mean-Field Bose Gas,
i.e. the model defined by the Hamiltonian
\begin{equation}
H_{\Lambda }^{MF}:=\stackunder{k\in \Lambda ^{*}}{\sum }\varepsilon
_{k}a_{k}^{*}a_{k}+\dfrac{\lambda _{0}}{2V}\left( N_{\Lambda
}^{2}-N_{\Lambda }\right) ,
\end{equation}
with no Bose condensations. Increase now the particle density. The
free-energy per particle, i.e., the chemical potential $\mu _{\beta ,\rho
}\leq \mu _{c},$ normally grows until we reach $\rho =\rho _{-}$. By further
increasing of the density, a Bose condensation continuously appears to reach
the value $x_{\rho _{+}}$ for $\rho =\rho _{+}.$ Meanwhile, the
corresponding chemical potential $\mu _{\rho }$ stays constant at the phase
transition: $\mu _{\rho }=\mu _{c}$ for $\rho \in [\rho _{-},\rho _{+}].$
Finally, at higher particle densities, i.e., for $\rho >\rho _{+}$, the Bose
condensate as well as the free-energy per particle $\mu _{\rho }>\mu _{c}$
both increase.

\section{Proofs: Large Deviations for a generalized Kac distribution\label%
{proof}}

We {are going to study the
grand-canonical ensemble at a fixed total particle density $\rho >0$}. But before doing this,
we start our analysis at a fixed chemical potential $\mu $. Then we prove {the LD principle
for the condensate plus ``out of condensate'' particle densities}. The corresponding
distribution $\Bbb{K}_{\Lambda ,\mu }$ is a combination of the so-called
\textit{Kac distribution} \cite{LewisPuleZagrebnov1} for particles outside
the condensate with the condensate distribution $\Bbb{D}_{\Lambda ,\mu }$.
This is {expressed by} Theorem \ref{theorem LDP-0}, which is therefore, {a
generalization of Theorem \ref{theorem LDP-0bis}}. To study the
phase transition, we use the \textit{generalized quasi-average procedure} \cite{LewisPuleZagrebnov1}
by taking a {"perturbed"} chemical potential
\begin{equation}
\tilde{\mu}_{c}:=\mu _{c}+\frac{\gamma }{\beta V}+o\left( \frac{1}{\beta V}%
\right) \mathrm{\ for\ }\gamma \in \Bbb{R},
\label{mu quasi-average procedure}
\end{equation}
we analyze the thermodynamic limit of the generalized Kac distribution at
this chemical potential, see Theorem \ref{theorem LDP-2}.

As a consequence, the generalized quasi-average procedure (\ref{mu
quasi-average procedure}) gives the finite volume behavior of the chemical
potential $\mu _{\Lambda ,\rho }$ solution of (\ref{mu fixed particle
density}) at the phase transition, i.e. when $\rho \in [\rho _{-},\rho _{+}]$
if $\rho _{+}>\rho _{-}.$ Indeed, by applying the distribution $\Bbb{K}%
_{\Lambda ,\mu }$ to an appropriate function, we obtain the mean particle
density at a chemical potential $\tilde{\mu}_{c}$ for any $\gamma \in \Bbb{R}
$. This procedure will then imply that for $\rho \in [\rho _{-},\rho _{+}]$
there is a unique and explicit $\gamma _{\rho }$ such that $\mu _{\Lambda
,\rho }=\tilde{\mu}_{c}$ with $|\gamma _{\rho }|=o(V)$, see Section \ref
{conclusion proof}.

Meanwhile, the large deviation principle for $\Bbb{K}_{\Lambda ,\mu }$ given
by Theorem \ref{theorem LDP-0} directly implies Theorem \ref{theorem
LDP-0bis} for any $\rho >0.$ Applying the result of Theorem \ref{theorem
LDP-2} to the chemical potential $\mu _{\Lambda ,\rho }=\tilde{\mu}_{c}$ for
$\gamma =\gamma _{\rho },$ we also get Theorem \ref{theorem LDP-1bisbis} for
$\rho \in [\rho _{-},\rho _{+}]$. If $\rho \notin [\rho _{-},\rho _{+}],$
the generalized quasi-average procedure is not necessary and Theorem \ref
{theorem LDP-1bis} is a simple consequence of Theorem \ref{theorem LDP-0bis}%
. We give now the promised proofs.

\subsection{Large deviations for generalized Kac distribution}

The particle number density as a $\Bbb{R}$-valued random variable, is
well-defined via a well-known probability measure, the so-called Kac
distribution \cite{LewisPuleZagrebnov1}. We give here a \textit{generalized} version
of the Kac distribution associated with the condensate and its depletion.
This distribution is defined, on the Borel subsets $\mathcal{A}\subset \Bbb{C%
}$ and $\mathcal{B}\subset \Bbb{R}_{+}$ by integration over the zero-mode coherent state:
\begin{equation}
\Bbb{K}_{\Lambda ,\mu }\left[ \mathcal{A}\right] :=e^{-\beta Vp_{\Lambda
}^{SB}(\beta ,\mu )}\frac{1}{2\pi }\stackunder{\mathcal{A}}{\int }\mathrm{d}%
^{2}c\stackunder{\mathcal{B}}{\int }\nu _{\Lambda }\left( \mathrm{d}y\right)
e^{\beta V\left( \mu \left( y+|c|^{2}\right) -f_{\Lambda }^{SB}\left( \beta
,y,c\right) \right) },
\end{equation}
with
\begin{equation}
\nu _{\Lambda }\left( \mathrm{d}y\right) :=\stackunder{n=1}{\stackrel{%
+\infty }{\sum }}\delta \left( \left[ yV\right] -n\right) \mathrm{d}y.
\label{definition of nu}
\end{equation}
Here $[.]$ is the integer part and
\begin{equation}
f_{\Lambda }^{SB}(\beta ,y,c):=-\dfrac{1}{\beta V}\ln \mathrm{Tr}_{\mathcal{H%
}_{B,k\neq 0}^{[yV]}}\left( \left\{ W_{\Lambda }\left( c\right) \right\}
^{([yV],k\neq 0)}\right) ,  \label{definition of the partial free energy}
\end{equation}
where $W_{\Lambda ,0}(c)$ results from the Bogoliubov approximation (\ref
{definition of Bogo approx}) of the statistical operator $W_{\Lambda ,0}$ (%
\ref{definition of the pressure}) and $\mathrm{A}^{(n,k\neq 0)}$ is the
restriction of any operator $\mathrm{A}\ $acting on the boson Fock space $%
\mathcal{F}_{B}^{\prime }$ to the space $\{\mathcal{H}_{0\Lambda }^{\bot
}\}^{\left( n\right) }$ of $n$ non-zero momentum bosons. Now we express our
first result concerning large deviations for the generalized Kac
distribution $\Bbb{K}_{\Lambda ,\mu }$.

\begin{theorem}[LD principle for the generalized Kac distribution]
\label{theorem LDP-0}\mbox{} \newline
The sequence $\{\Bbb{K}_{\Lambda ,\mu }\}$ satisfies a large deviation
principle with speed\textit{\ }$\beta V$ and rate function
\[
\mathrm{K}_{\mu }\left( x,y\right) :=p^{SB}\left( \beta ,\mu \right)
+f_{0}^{B}\left( \beta ,y,x\right) +\frac{\lambda _{0}}{2}\left( y+x\right)
^{2}-\mu \left( y+x\right) .
\]
Here $x=|c|^{2}\geq 0,$ $y\geq 0$ and
\[
f_{0}^{B}\left( \beta ,y,x\right) :=\stackunder{\alpha \leq 0}{\sup }\left\{
\alpha (y+x)-p_{0}^{B}\left( \beta ,\alpha ,x\right) \right\}
\]
is the Legendre-Fenchel transform of $p_{0}^{B}\left( \beta ,\alpha
,x\right) $ (\ref{eq 4.26}).
\end{theorem}

\noindent \textit{Proof}. Let us start by some observations. The pressure $%
p_{0}^{B}(\beta ,\alpha ,x)$ defined in (\ref{eq 4.26}) and used in (\ref
{pressure fixed density_1}) can be explicitly computed.\ Indeed,
\begin{eqnarray}
p_{0}^{B}\left( \beta ,\alpha ,x\right)  &=&\alpha x-\dfrac{1}{\beta \left(
2\pi \right) ^{3}}\stackunder{\Bbb{R}^{3}}{\dint }\ln \left( 1-e^{-\beta
\sqrt{\left( \varepsilon _{k}-\alpha \right) \left( \varepsilon _{k}-\alpha
+2x\lambda _{k}\right) }}\right) \mathrm{d}^{3}k+  \nonumber \\
&&+\dfrac{1}{2\left( 2\pi \right) ^{3}}\stackunder{\Bbb{R}^{3}}{\dint }%
\left\{ \varepsilon _{k}-\alpha +x\lambda _{k}-\sqrt{\left( \varepsilon
_{k}-\alpha \right) \left( \varepsilon _{k}-\alpha +2x\lambda _{k}\right) }%
\right\} \mathrm{d}^{3}k,  \label{eq 4.26bis}
\end{eqnarray}
for any $\alpha \leq 0$. Since $p_{0}^{B}\left( \beta ,\alpha ,x\right) $ is
a convex function of $\alpha \leq 0$, it is also the Legendre-Fenchel
transform of $f_{0}^{B}\left( \beta ,y,x\right) ,$ i.e.
\begin{equation}
p_{0}^{B}\left( \beta ,\alpha ,x\right) =\stackunder{y\geq 0}{\sup }\left\{
\alpha (y+x)-f_{0}^{B}\left( \beta ,y,x\right) \right\} \mathrm{\ for\ any\;}%
\alpha \leq 0.
\end{equation}
Combined with (\ref{pressure fixed density_1}) this last inequality implies
that
\begin{equation}
p^{SB}\left( \beta ,\mu \right) =\stackunder{x\geq 0}{\sup }\left\{
\stackunder{\alpha \leq 0}{\inf }\left\{ \stackunder{y\geq 0}{\sup }\left\{
\alpha (y+x)-f_{0}^{B}\left( \beta ,y,x\right) +\frac{(\mu -\alpha )^{2}}{%
2\lambda _{0}}\right\} \right\} \right\} .
\label{pressure SB as function of y-x}
\end{equation}
We would like to bring the infimum over $\alpha \leq 0$ inside the two other
suprema. In general, a supremum and an infimum do not commute. In this
peculiar case, this is however the case. Indeed, for any fixed $x\geq 0$ the
function
\begin{equation}
\Psi \left( y,\alpha \right) :=\alpha (y+x)-f_{0}^{B}\left( \beta
,y,x\right) +\frac{\left( \mu -\alpha \right) ^{2}}{2\lambda _{0}}
\end{equation}
is a strictly concave function of $y\geq 0$ and a strictly convex function
of $\alpha \leq 0$. Then, we obtain the uniqueness of the stationary point $(%
\tilde{y},\tilde{\alpha})$ solution of
\begin{equation}
\partial _{y}\Psi \left( y,\alpha \right) =0\mathrm{\ and\;}\partial
_{\alpha }\Psi \left( y,\alpha \right) =y+x+\frac{\alpha -\mu }{\lambda _{0}}%
=0.
\end{equation}
In particular, we can commute the infimum over $\alpha \leq 0$ and the
supremum over $y\geq 0$ in (\ref{pressure SB as function of y-x}) to obtain
\begin{equation}
p^{SB}\left( \beta ,\mu \right) =\stackunder{(x,y)\in \Bbb{R}_{+}^{2}}{\sup }%
\left\{ \mu \left( y+x\right) -f_{0}^{B}\left( \beta ,y,x\right) -\frac{%
\lambda _{0}}{2}\left( y+x\right) ^{2}\right\} .
\label{pressure SB as function of y-xbis}
\end{equation}
This result is coherent with the rate function $\mathrm{K}_{\mu }\left(
x,y\right) $. By explicit computations, observe also that there are $M,B>0$
such that any solution $(x_{\mu },y_{\mu })$ of the variational problem (\ref
{pressure SB as function of y-xbis}) verifies $x_{\mu }<M$ and $y_{\mu }<M$
whereas for any $x\geq M$ and $y\geq M$ we have
\begin{equation}
\mu \left( y+x\right) -f_{0}^{B}\left( \beta ,y,x\right) -\frac{\lambda _{0}%
}{2}\left( y+x\right) ^{2}\leq -B\left({y + x}\right) .
\label{superstability consequence}
\end{equation}
Now we are in position to analyze the LD principle for
distribution $\Bbb{K}_{\Lambda ,\mu }$ (Section \ref{Section LDP}).

From (\ref{eq 4.26bis}) the rate function $\mathrm{K}_{\mu }\left(
x,y\right) $ is not identical $\infty $ and has compact level sets, i.e. for
each $m<\infty $, the subset $\left\{ \left( x,y\right) :\mathrm{K}_{\mu
}\left( x,y\right) \leq m\right\} $ is compact.

Let a closed set $\mathcal{C}:=\mathcal{C}_{0}\times \mathcal{C}_{1}\subset
\Bbb{C\times R}_{+}.$ Remark that $M$ can be taken arbitrary large (and $B$
being the same). Then, without lost of generality, we can assume that any $%
c\in \mathcal{C}_{0}$ and $y\in \mathcal{C}_{1}$ satisfy $|c|^{2}<M$ and $y<M
$ respectively. By (\ref{superstability consequence}), we also obtain
\begin{eqnarray}
\Bbb{K}_{\Lambda ,\mu }\left[ \mathcal{C}\right]  &\leq &\frac{1}{2\pi }%
e^{\beta V\left\{ \stackunder{\mathcal{C}}{\sup }\left\{ \mu
(y+|c|^{2})-f_{\Lambda }^{SB}(\beta ,y,c)\right\} -p_{\Lambda }^{SB}\left(
\beta ,\mu \right) \right\} }\stackunder{\mathcal{C}_{0}}{\int }\mathrm{d}%
^{2}c\stackunder{\mathcal{C}_{1}}{\int }\nu _{\Lambda }\left( \mathrm{d}%
y\right) +  \nonumber \\
&&+\frac{1}{2\pi }e^{-\beta V\left\{ p_{\Lambda }^{SB}\left( \beta ,\mu
\right) +2BM\right\} }\stackunder{\Bbb{C}}{\int }\mathrm{d}^{2}c\stackunder{%
\Bbb{R}_{+}}{\int }\nu _{\Lambda }\left( \mathrm{d}y\right) e^{-\beta
VB\left( |c|^{2}+y\right) }.  \label{superstability consequencebis}
\end{eqnarray}
For large enough $M,$ one has
\begin{equation}
2BM+\stackunder{\mathcal{C}}{\sup }\left\{ \mu \left( y+x\right)
-f_{0}^{B}\left( \beta ,y,x\right) -\frac{\lambda _{0}}{2}\left( y+x\right)
^{2}\right\} >0.
\end{equation}
Consequently, the inequality (\ref{superstability consequencebis}) combined
Lemma \ref{LDP-lemma1bis} implies that
\begin{equation}
\stackunder{\Lambda }{\lim \sup }\frac{1}{\beta V}\ln \Bbb{K}_{\Lambda ,\mu
}\left[ \mathcal{C}\right] \leq -\stackunder{\mathcal{C}}{\inf }\mathrm{K}%
_{\mu }\left( |c|^{2},y\right) .
\end{equation}
In other words, the large deviations upper bound (\ref{LDPupper}) for $\Bbb{K%
}_{\Lambda ,\mu }$ with speed $\beta V$ and rate function $\mathrm{K}_{\mu }$
is verified. It remains to analyze the corresponding large deviations lower
bound (\ref{LDPlower}).

Let $\mathcal{G}$ be an arbitrary open subset of $\Bbb{C\times R}_{+}$. Note
that
\begin{equation}
\Bbb{K}_{\Lambda ,\mu }\left[ \mathcal{G}\right] \geq \Bbb{K}_{\Lambda ,\mu
}\left[ \left\{ (c,y)\right\} \right] =e^{-\beta Vp_{\Lambda }^{SB}\left(
\beta ,\mu \right) }e^{\beta V\left\{ \mu ([yV]+|c|^{2})-f_{\Lambda
}^{SB}(\beta ,y,c)\right\} },
\end{equation}
with $(c,y)\in \mathcal{G}$. From Lemma \ref{LDP-lemma1bis}, it yields that
\begin{equation}
\stackunder{\Lambda }{\lim \inf }\frac{1}{\beta V}\ln \Bbb{K}_{\Lambda ,\mu
}\left[ \mathcal{G}\right] \geq -\mathrm{K}_{\mu }\left( |c|^{2},y\right) .
\end{equation}
Since the last inequality holds for each point of $\mathcal{G}$, it means
that
\begin{equation}
\stackunder{\Lambda }{\lim \inf }\frac{1}{\beta V}\ln \Bbb{K}_{\Lambda ,\mu
}\left[ \mathcal{G}\right] \geq -\stackunder{\mathcal{G}}{\inf }\mathrm{K}%
_{\mu }\left( |c|^{2},y\right) ,
\end{equation}
i.e. the corresponding large deviation lower bound (\ref{LDPlower}) for $%
\Bbb{K}_{\Lambda ,\mu }$ holds with speed\textit{\ }$\beta V$ and rate
function $\mathrm{K}_{\mu }$. $\square $

For $\mu \neq \mu _{c}$ we already know \cite{Adams-Bru1bis} that the
variational problem (\ref{pressure SB as function of y-xbis}) has a unique
solution $(x_{\mu },y_{\mu })$. Therefore, as a direct consequence of the
fact that the sequence $\{\Bbb{K}_{\Lambda ,\mu }\}$ satisfies a large
deviations principle with rate function $\mathrm{K}_{\mu }$ having a unique
minimum in $\Bbb{R}_{+}^{2}$ at $(x_{\mu },y_{\mu })$ for any $\mu \neq \mu
_{c}$, the distribution $\Bbb{K}_{\Lambda ,\mu }$ converges weakly on the
set of probability measures $\mathcal{M}_{1}\left( \Bbb{C\times R}%
_{+}\right) $ as $\Lambda \uparrow \Bbb{R}^{3}$ towards the singular measure
\begin{equation}
\Bbb{K}_{\mu }:=\stackunder{\Lambda }{\lim }\Bbb{K}_{\Lambda ,\mu }=\frac{1}{%
2\pi }\stackunder{0}{\stackrel{2\pi }{\int }}\delta \left( c-x_{\mu
}^{1/2}e^{i\theta }\right) \delta \left( y-y_{\mu }\right) \mathrm{d}\theta ,%
\mathrm{\ for\ }\mu \neq \mu _{c}.  \label{equation 1}
\end{equation}
Now, the next step is to evaluate the limiting probability measure at the
phase transition defined for a chemical potential $\mu =\mu _{c}$. Indeed,
if $\rho _{+}>\rho _{-},$ the solution $(x_{\mu },y_{\mu })$ jumps when $\mu
$ cross the critical chemical potential $\mu _{c}$ from $(0,\rho _{-})$ to $%
(x_{\rho _{+}},y_{\rho _{+}})$ with $x_{\rho _{+}}>0$ and $y_{\rho
_{+}}:=\rho _{+}-x_{\rho _{+}}>\rho _{-}.$

\begin{theorem}[The generalized Kac distribution at the phase transition]
\label{theorem LDP-2}\mbox{} \newline
If $\rho _{+}>\rho _{-},$ then the distribution $\Bbb{K}_{\Lambda ,\tilde{\mu%
}_{c}}$ converges weakly in $\mathcal{M}_{1}\left( \Bbb{C\times R}%
_{+}\right) $ as $\Lambda \uparrow \Bbb{R}^{3}$ towards
\begin{equation}
\stackunder{\Lambda }{\lim }\Bbb{K}_{\Lambda ,\tilde{\mu}_{c}}=\xi _{\gamma
}\delta \left( c\right) \delta \left( y-\rho _{-}\right) +\frac{\left( 1-\xi
_{\gamma }\right) }{2\pi }\stackunder{0}{\stackrel{2\pi }{\int }}\delta
\left( c-x_{\rho _{+}}^{1/2}e^{i\theta }\right) \delta \left( y-y_{\rho
_{+}}\right) \mathrm{d}\theta ,  \label{theorem LDP-2bis}
\end{equation}
with $\xi _{\gamma }:=(1+e^{\gamma (\rho _{+}-\rho _{-})})^{-1}\in (0,1)$
and $\tilde{\mu}_{c}$ defined by (\ref{mu quasi-average procedure}) for any $%
\gamma \in \Bbb{R}$.
\end{theorem}

\noindent \textit{Proof. }We have already mentioned that the rate function $%
\mathrm{K}_{\mu _{c}}$ has two distinct minima in $\Bbb{R}_{+}^{2}$ at $%
(0,\rho _{-})$ and $(x_{\rho _{+}},y_{\rho _{+}})$. To get around this
complication, take $\varepsilon \in (0,x_{\rho _{+}})\cap (0,y_{\rho
_{+}}-\rho _{-})$ and define
\begin{equation}
\mathcal{A}_{-}:=\left\{ c\in \Bbb{C}:|c|^{2}\in \left( 0,x_{\rho
_{+}}-\varepsilon \right] \right\} \times \left( \rho _{-},y_{\rho
_{+}}-\varepsilon \right]
\end{equation}
and
\begin{equation}
\mathcal{A}_{+}:=\left\{ c\in \Bbb{C}:|c|^{2}\in \left( x_{\rho
_{+}}-\varepsilon ,+\infty \right) \right\} \times \left( y_{\rho
_{+}}-\varepsilon ,+\infty \right) .
\end{equation}
Now, let $\mathrm{K}_{\mu _{c}^{-}}$ and $\mathrm{K}_{\mu _{c}^{+}}$ be
defined as the two restrictions of $\mathrm{K}_{\mu _{c}}$ to $\mathcal{A}%
_{-}$ and $\mathcal{A}_{+}$ respectively and remark that $\mathrm{K}_{\mu
_{c}^{-}}$ and $\mathrm{K}_{\mu _{c}^{+}}$ have both a unique minimizer in $%
\Bbb{R}_{+}^{2}$, respectively $(0,\rho _{-})$ and $(x_{\rho _{+}},y_{\rho
_{+}})$. Define the corresponding probability measures
\begin{equation}
\Bbb{L}_{\Lambda }^{-}\left[ \mathcal{A}\right] :=\frac{\Bbb{K}_{\Lambda
,\mu _{c}}\left[ \mathcal{A}\cap \mathcal{A}_{-}\right] }{\Bbb{K}_{\Lambda
,\mu _{c}}\left[ \mathcal{A}_{-}\right] }\mathrm{\ and\ }\Bbb{L}_{\Lambda
}^{+}\left[ \mathcal{A}\right] :=\frac{\Bbb{K}_{\Lambda ,\mu _{c}}\left[
\mathcal{A}\cap \mathcal{A}_{+}\right] }{\Bbb{K}_{\Lambda ,\mu _{c}}\left[
\mathcal{A}_{+}\right] },
\end{equation}
which satisfy a large deviations principle respectively\textit{\ }with rate
functions $\mathrm{K}_{\mu _{c}^{-}}$ and $\mathrm{K}_{\mu _{c}^{+}}$ (where
$x=|c|^{2}$). Take any positive and continuous function $\varphi (c,y)$ of $%
(c,y)\in \Bbb{C}\times \Bbb{R}_{+}$ and observe that
\begin{equation}
\stackunder{\Bbb{R}_{+}^{2}}{\int }\varphi \left( c,y\right) \Bbb{K}%
_{\Lambda ,\mu _{c,\Lambda }}\left( \mathrm{d}^{2}c\right) =\frac{%
\stackunder{\Bbb{C}}{\int }\mathrm{d}^{2}c\stackunder{\Bbb{R}_{+}}{\int }\nu
_{\Lambda }\left( \mathrm{d}y\right) \varphi \left( c,y\right) e^{\beta
V\left( \tilde{\mu}_{c}(y+|c|^{2})-f_{\Lambda }^{SB}(\beta ,y,c)\right) }}{%
\stackunder{\Bbb{C}}{\int }\mathrm{d}^{2}c\stackunder{\Bbb{R}_{+}}{\int }\nu
_{\Lambda }\left( \mathrm{d}y\right) e^{\beta V\left( \tilde{\mu}%
_{c}(y+|c|^{2})-f_{\Lambda }^{SB}(\beta ,y,c)\right) }}=\Phi _{\Lambda
}^{-}+\Phi _{\Lambda }^{+},
\end{equation}
with
\begin{eqnarray}
\Phi _{\Lambda }^{-} &:&=\dfrac{\stackunder{\mathcal{A}_{-}}{\int }\varphi
\left( c,y\right) e^{\left\{ \gamma +o(1)\right\} (y+|c|^{2})}\Bbb{L}%
_{\Lambda }^{-}\left( \mathrm{d}^{2}c\mathrm{d}y\right) }{\stackunder{%
\mathcal{A}_{-}}{\int }e^{\left\{ \gamma +o(1)\right\} (y+|c|^{2})}\Bbb{L}%
_{\Lambda }^{-}\left( \mathrm{d}^{2}c\mathrm{d}y\right) +\Theta _{\Lambda }%
\stackunder{\mathcal{A}_{+}}{\int }e^{\left\{ \gamma +o(1)\right\}
(y+|c|^{2})}\Bbb{L}_{\Lambda }^{+}\left( \mathrm{d}^{2}c\mathrm{d}y\right) },
\\
\Phi _{\Lambda }^{+} &:&=\dfrac{\stackunder{\mathcal{A}_{+}}{\int }\varphi
\left( c,y\right) e^{\left\{ \gamma +o(1)\right\} (y+|c|^{2})}\Bbb{L}%
_{\Lambda }^{+}\left( \mathrm{d}^{2}c\mathrm{d}y\right) }{\Theta _{\Lambda
}^{-1}\stackunder{\mathcal{A}_{-}}{\int }e^{\left\{ \gamma +o(1)\right\}
(y+|c|^{2})}\Bbb{L}_{\Lambda }^{-}\left( \mathrm{d}^{2}c\mathrm{d}y\right) +%
\stackunder{\mathcal{A}_{+}}{\int }e^{\left\{ \gamma +o(1)\right\}
(y+|c|^{2})}\Bbb{L}_{\Lambda }^{+}\left( \mathrm{d}^{2}c\mathrm{d}y\right) },
\end{eqnarray}
and
\begin{equation}
\Theta _{\Lambda }:=\dfrac{\stackunder{\mathcal{A}_{+}}{\int }e^{\beta
V\left\{ \mu _{c}(y+|c|^{2})-f_{\Lambda }^{SB}(\beta ,y,c)\right\} }\nu
_{\Lambda }\left( \mathrm{d}y\right) \mathrm{d}^{2}c}{\stackunder{\mathcal{A}%
_{-}}{\int }e^{\beta V\left\{ \mu _{c}(y+|c|^{2})-f_{\Lambda }^{SB}(\beta
,y,c)\right\} }\nu _{\Lambda }\left( \mathrm{d}y\right) \mathrm{d}^{2}c}.
\label{coefficient theta}
\end{equation}
By Lemma \ref{LDP-lemma1bis} the function
\begin{equation}
\mu _{c}(y+|c|^{2})-f_{\Lambda }^{SB}(\beta ,y,c)
\end{equation}
converges in the thermodynamic limit to
\begin{equation}
\mu _{c}(y+|c|^{2})-f_{0}^{B}\left( \beta ,y,|c|^{2}\right) -\frac{\lambda
_{0}}{2}\left( y+|c|^{2}\right) ^{2},
\end{equation}
which has suprema at $(0,\rho _{-})$ and $(e^{i\theta }x_{\rho _{+}},y_{\rho
_{+}})$ for any $\theta \in [0,2\pi ].$ Consequently, the coefficient $%
\Theta _{\Lambda }$ (\ref{coefficient theta}) converges to $1$ in the
thermodynamic limit. Since $\rho _{+}=y_{\rho _{+}}+x_{\rho _{+}}$, it is
then straightforward to see that
\begin{equation}
\stackunder{\Lambda }{\lim }\Phi _{\Lambda }^{-}=\xi _{\gamma }\varphi
(0,\rho _{-})\mathrm{\ and\ }\stackunder{\Lambda }{\lim }\Phi _{\Lambda
}^{+}=\frac{(1-\xi _{\gamma })}{2\pi }\stackunder{0}{\stackrel{2\pi }{\int }}%
\varphi (x_{\rho _{+}}^{1/2}e^{i\theta },y_{\rho _{+}})\mathrm{d}\theta .
\end{equation}
{Let us apply} these limits to the function $\varphi (c,y)=e^{-t(c+y)}$ with $t>0.$
Then, by bijectivity of the Laplace transform, it follows that $\Bbb{K}%
_{\Lambda ,\tilde{\mu}_{c}}$ converges weakly on $\mathcal{M}_{1}\left( \Bbb{%
C\times R}_{+}\right) $ as $\Lambda \uparrow \Bbb{R}^{3}$ to (\ref
{theorem LDP-2bis}).$\;\square $

Notice that the function $\xi _{\gamma }:\Bbb{R}\rightarrow \left(
0,1\right) $ defined in Theorem \ref{theorem LDP-2} is strictly decreasing
and in fact bijective. Therefore, by applying $\Bbb{K}_{\Lambda ,\tilde{\mu}%
_{c}}$ to $\varphi (c,y)=|c|^{2}+y,$ we have shown that the particle density
can converge to any fixed density in the open set $\left( \rho _{-},\rho
_{+}\right) :$%
\begin{equation}
\stackunder{\Lambda }{\lim }\left\langle \frac{N_{\Lambda }}{V}\right\rangle
_{H_{\Lambda ,\lambda _{0}}^{SB}}=\xi _{\gamma }\rho _{-}+\left( 1-\xi
_{\gamma }\right) \rho _{+}.  \label{particle density phase transition}
\end{equation}
Note that all these results are coherent since we have
\begin{equation}
\rho _{-}=\stackunder{\gamma \rightarrow -\infty }{\lim }\left\{ \xi
_{\gamma }\rho _{-}+\left( 1-\xi _{\gamma }\rho _{+}\right) \right\} \mathrm{%
\ and\;}\rho _{+}=\stackunder{\gamma \rightarrow +\infty }{\lim }\left\{ \xi
_{\gamma }\rho _{-}+\left( 1-\xi _{\gamma }\rho _{+}\right) \right\} .
\label{particle density limiting case}
\end{equation}
In particular, if $\gamma =\gamma _{\Lambda }=o\left( \pm V\right) $ in (\ref
{mu quasi-average procedure}) diverges to $\pm \infty $, then we would
obtain one of the previous limit, depending if $\gamma _{\Lambda }\downarrow
-\infty $ or $\gamma _{\Lambda }\uparrow +\infty .$

\subsection{Application of the generalized Kac distribution for a fixed
particle density\label{conclusion proof}}

Let us consider now {the total particle density as a parameter that defines} the grand-canonical
ensemble. Theorems \ref{theorem LDP-0bis} and \ref{theorem LDP-1bisbis} are
direct consequences respectively of Theorem \ref{theorem LDP-0} and (\ref
{equation 1}) for the chemical potential $\mu _{\rho }$ defined as the
thermodynamic limit of $\mu _{\Lambda ,\rho }$ (\ref{mu fixed particle
density}). The only {remaining question is to study the case of fixed particle densities at
the the point of phase transition, i.e. in  domain:} $\rho \in \left( \rho _{-},\rho _{+}\right) $
for $\rho _{+}>\rho _{-}.$ From (\ref{particle density phase transition}), we
obtain that
\begin{equation}
\stackunder{\Lambda }{\lim }\left\langle \frac{N_{\Lambda }}{V}\right\rangle
_{H_{\Lambda ,\lambda _{0}}^{SB}}=\rho \in \left( \rho _{-},\rho _{+}\right)
,\text{ }
\end{equation}
for a chemical potential
\begin{equation}
\mu =\mu _{c}+\frac{\gamma _{\rho }}{\beta V}+o\left( \frac{1}{\beta V}%
\right) \text{\textrm{\ with }}\gamma _{\rho }:=\frac{1}{\rho _{+}-\rho _{-}}%
\ln \left( \frac{\rho -\rho _{-}}{\rho _{+}-\rho }\right) ,
\label{explicit gamma}
\end{equation}
cf. (\ref{mu quasi-average procedure}). Therefore,
\begin{equation}
\mu _{\Lambda ,\rho }=\mu _{c}+\frac{\gamma _{\rho }}{\beta V}+o\left( \frac{%
1}{\beta V}\right) .
\end{equation}
In particular, from (\ref{theorem LDP-2}) with $\gamma =\gamma _{\rho }\ $we
get Theorem \ref{theorem LDP-1bisbis} for $\rho \in \left( \rho _{-},\rho
_{+}\right) $. Recall also (\ref{particle density limiting case}). In other
words, if $\rho =\rho _{-}$ then $\gamma _{\rho }<0$ ($|\gamma _{\rho }|=o(V)
$) would diverges to $-\infty ,$ whereas if $\rho =\rho _{+}$ then $\gamma
_{\rho }=o\left( V\right) \rightarrow +\infty $. It follows that Theorem \ref
{theorem LDP-1bisbis} is proven for any $\rho \in [\rho _{-},\rho _{+}].$

\section{Appendix\label{section appendix}}

In this appendix, we first give supplementary results needed in the previous
section. Next, for the convenience of our reader, we shortly repeat the
notion of large deviations principles.

\subsection{Some technical statements and proofs}

The thermodynamic limit of $p_{\Lambda }^{SB}(\beta ,\mu ,c)$ (\ref
{definition of the pressure as function of c}) is first analyzed in order to
obtain next the one of the free-energy density $f_{\Lambda }^{SB}(\beta ,y,c)
$ (\ref{definition of the partial free energy}), which is given in Lemma \ref
{LDP-lemma1bis}.

\begin{lemma}[The pressure $p_{\Lambda }^{SB}(\beta ,\mu ;c)$ in the thermodynamic limit]
\label{LDP-lemma1}\mbox{} \newline
For any $c\in \Bbb{C}$, $\mu \in \Bbb{R}$ and $\beta >0$, the pressure $%
p_{\Lambda }^{SB}(\beta ,\mu ,c)$ converges towards
\[
p^{SB}(\beta ,\mu ,c):=\stackunder{\Lambda }{\lim }p_{\Lambda }^{SB}(\beta
,\mu ,c)=\stackunder{\alpha \leq 0}{\inf }\left\{ p_{0}^{B}(\beta ,\alpha
,x)+\frac{(\mu -\alpha )^{2}}{2\lambda _{0}}\right\} .
\]
Here $x=|c|^{2}\geq 0$ and recall that $p_{0}^{B}(\beta ,\alpha ,x)$ is
defined in (\ref{eq 4.26}), cf.\ also (\ref{eq 4.26bis}).
\end{lemma}

\noindent \textit{Proof.} The proof is obtained by a comparison between
suitable lower and upper bounds for $p_{\Lambda }^{SB}(\beta ,\mu ,c).$ We
start by the lower bound. By taking any othonormal basis $\{\langle \psi
_{n}^{\prime }|\}_{n=1}^{\infty }$ of $\mathcal{F}_{\Lambda }^{\prime }$,
\begin{equation}
\mathrm{Tr}_{\mathcal{F}_{\Lambda }^{\prime }}\left\{ W_{\Lambda }\left(
c\right) \right\} =\stackunder{n=1}{\stackrel{\infty }{\sum }}\left\langle
c\otimes \psi _{n}^{\prime }\right| e^{-\beta (H_{\Lambda ,\lambda
_{0}}^{SB}-\mu N_{\Lambda })}\left| c\otimes \psi _{n}^{\prime
}\right\rangle ,
\end{equation}
and so, by the Peierls-Bogoliubov inequality we get
\begin{equation}
\mathrm{Tr}_{\mathcal{F}_{\Lambda }^{\prime }}\left\{ W_{\Lambda }\left(
c\right) \right\} \geq \stackunder{\left\{ \psi _{n}^{\prime }\right\}
_{n=1}^{\infty }}{\sup }\left\{ \stackunder{n=1}{\stackrel{\infty }{\sum }}%
e^{-\beta \left\langle c\otimes \psi _{n}^{\prime }\right| H_{\Lambda
,\lambda _{0}}^{SB}-\mu N_{\Lambda }\left| c\otimes \psi _{n}^{\prime
}\right\rangle }\right\} =\mathrm{Tr}_{\mathcal{F}_{\Lambda }^{\prime
}}\left\{ e^{-\beta H_{\Lambda ,\lambda _{0}}^{SB}\left( c,\mu \right)
}\right\} ,  \label{eq cond2demo2}
\end{equation}
see e.g. \cite{ReedSimon,Simon1}, where $H_{\Lambda ,\lambda
_{0}}^{SB}(c,\mu )$ results from the Bogoliubov approximation (\ref
{definition of Bogo approx}) of $\{H_{\Lambda ,\lambda _{0}}^{SB}-\mu
N_{\Lambda }\}.$ From \cite{Adams-Bru1bis} we already know that
\begin{equation}
\stackunder{\Lambda }{\lim }\frac{1}{\beta V}\ln \mathrm{Tr}_{\mathcal{F}%
_{\Lambda }^{\prime }}\left\{ e^{-\beta H_{\Lambda ,\lambda _{0}}^{SB}\left(
c,\mu \right) }\right\} =\stackunder{\alpha \leq 0}{\inf }\left\{
p_{0}^{B}(\beta ,\alpha ,|c|^{2})+\frac{(\mu -\alpha )^{2}}{2\lambda _{0}}%
\right\} .  \label{limit sup 1}
\end{equation}
Consequently, the inequality (\ref{eq cond2demo2}) implies in the
thermodynamic limit the lower bound
\begin{equation}
p^{SB}(\beta ,\mu ,c)\geq \stackunder{\alpha \leq 0}{\inf }\left\{
p_{0}^{B}(\beta ,\alpha ,|c|^{2})+\frac{(\mu -\alpha )^{2}}{2\lambda _{0}}%
\right\} ,  \label{pressure avec c lower bound}
\end{equation}
for any $c\in \Bbb{C}$, $\mu \in \Bbb{R}$ and $\beta >0.$

{To obtain an upper bound on $p^{SB}(\beta ,\mu ,c)$, we follow the idea of  \cite
{LiebSeiringerYngvason3}, and use the coherent state} representation of $\{H_{\Lambda
,\lambda _{0}}^{SB}-\mu N_{\Lambda }\}$ given by
\begin{equation}
H_{\Lambda ,\lambda _{0}}^{SB}-\mu N_{\Lambda }=\stackunder{\Bbb{C}}{\int }%
\mathrm{d}^{2}c\left\{ \hat{H}_{\Lambda ,\lambda _{0}}^{SB}\left( c,\mu
\right) \left| c\right\rangle \left\langle c\right| \right\} ,
\end{equation}
where the Hamiltonian $\hat{H}_{\Lambda ,\lambda _{0}}^{SB}\left( c,\mu
\right) $ is defined on $\mathcal{F}_{\Lambda }^{\prime }$ by
\begin{equation}
\hat{H}_{\Lambda ,\lambda _{0}}^{SB}\left( c,\mu \right) :=H_{\Lambda
,\lambda _{0}}^{SB}\left( c,\mu \right) +\Delta ,
\end{equation}
with
\begin{equation}
\Delta :=\mu -2\lambda _{0}|c|^{2}+\frac{\lambda _{0}}{V}-\frac{1}{V}%
\stackunder{k\in \Lambda ^{*}\backslash \left\{ 0\right\} }{\sum }\left(
\lambda _{0}+\lambda _{k}\right) a_{k}^{*}a_{k}.
\label{pressure avec c upper bound-2}
\end{equation}
Actually, $\hat{H}_{\Lambda ,\lambda _{0}}^{SB}\left( c,\mu \right) $ is
derived by replacing the operators $a_{0}^{*}a_{0}$, $a_{0}a_{0}$, $%
a_{0}^{*}a_{0}^{*}$, and $a_{0}^{*}a_{0}^{*}a_{0}a_{0}$ in $\{H_{\Lambda
,\lambda _{0}}^{SB}-\mu N_{\Lambda }\}$ respectively by $|Vc|^{2}-1,$ $%
Vc^{2},$ $V\bar{c}^{2}$ and $V^{2}|c|^{4}-4V|c|^{2}+2.$ Let $\{\langle \psi
_{n}^{\prime }\left( c\right) |\}_{n=1}^{\infty }$ be an othonormal basis of
eigenvectors of $\hat{H}_{\Lambda ,\lambda _{0}}^{SB}\left( c,\mu \right) .$
Since for any $z,c\in \Bbb{C}$
\begin{equation}
\left\langle z|c\right\rangle =e^{-\frac{1}{2}\left\{ \left( \bar{z}-\bar{c}%
\right) \left( z-c\right) +\bar{c}z-\bar{z}c\right\} },
\end{equation}
it follows that
\begin{eqnarray}
\mathrm{Tr}_{\mathcal{F}_{\Lambda }^{\prime }}\left\{ W_{\Lambda }\left(
c\right) \right\}  &=&\stackunder{n=1}{\stackrel{\infty }{\sum }}%
\left\langle c\otimes \psi _{n}^{\prime }\left( c\right) \right| e^{-\beta
\stackunder{\Bbb{C}}{\int }\mathrm{d}^{2}z\hat{H}_{\Lambda ,\lambda
_{0}}^{SB}\left( z,\mu \right) \left| z\right\rangle \left\langle z\right|
}\left| c\otimes \psi _{n}^{\prime }\left( c\right) \right\rangle   \nonumber
\\
&=&\stackunder{n=1}{\stackrel{\infty }{\sum }}\left\{ 1+\stackunder{m=1}{%
\stackrel{\infty }{\sum }}\frac{\left( -\beta \right) ^{m}}{m!}\stackunder{%
\Bbb{C}^{m}}{\int }\mathrm{d}^{2}z_{1}\cdots \mathrm{d}^{2}z_{m}e^{-\frac{V}{%
2}\left\{ \mathrm{R}_{m}\left( z_{1},\cdots ,z_{m}\right) +i\mathrm{I}%
_{m}\left( z_{1},\cdots ,z_{m}\right) \right\} }\right.   \nonumber \\
&&\left. \times \stackunder{j=1}{\stackrel{m}{\prod }}\left\langle \psi
_{n}^{\prime }\left( c\right) \right| \hat{H}_{\Lambda ,\lambda
_{0}}^{SB}\left( z_{j},\mu \right) \left| \psi _{n}^{\prime }\left( c\right)
\right\rangle \right\} ,  \label{important inequality0}
\end{eqnarray}
with the two real-valued functions $\mathrm{R}_{m}$ and $\mathrm{I}_{m}$ of $%
\left( z_{1},\cdots ,z_{m}\right) \in \Bbb{C}^{m}$ defined by
\begin{equation}
\begin{array}{l}
\mathrm{R}_{m}\left( z_{1},\cdots ,z_{m}\right) :=\left| z_{1}-c\right| ^{2}+%
\stackunder{j=1}{\stackrel{m}{\sum }}\left| z_{j-1}-z_{j}\right| ^{2}+\left|
z_{m}-c\right| ^{2}, \\
\mathrm{I}_{m}\left( z_{1},\cdots ,z_{m}\right) :=i\left( \bar{z}_{1}c-\bar{c%
}z_{1}\right) +i\stackunder{j=1}{\stackrel{m}{\sum }}\left( \bar{z}%
_{j}z_{j-1}-\bar{z}_{j-1}z_{j}\right) +i\left( \bar{c}z_{m}-\bar{z}%
_{m}c\right) .
\end{array}
\end{equation}
Since $\mathrm{I}_{m}\left( c,\cdots ,c\right) =0$ and
\begin{equation}
\stackunder{\left( z_{1},\cdots ,z_{m}\right) \in \Bbb{C}^{m}}{\inf }\mathrm{%
R}_{m}\left( z_{1},\cdots ,z_{m}\right) =\mathrm{R}_{m}\left( c,\cdots
,c\right) =0,
\end{equation}
{by virtue of} (\ref{important inequality0}) combined with large deviations arguments,
one can obtain in the thermodynamic limit that
\begin{equation}
p^{SB}(\beta ,\mu ,c)=\stackunder{\Lambda }{\lim }\frac{1}{\beta V}\ln
\mathrm{Tr}_{\mathcal{F}_{\Lambda }^{\prime }}\left\{ e^{-\beta \hat{H}%
_{\Lambda ,\lambda _{0}}^{SB}\left( c,\mu \right) }\right\} .
\label{important inequality}
\end{equation}
{Justification of the LD technique in sums (\ref{important inequality0}) is based on the uniform domination theorem
and it follows the line of reasoning developed in
\cite{LiebSeiringerYngvason3}.}
Meanwhile, by using the Bogoliubov convexity inequality \cite{BruZagrebnov8}
it follows that
\begin{equation}
\mathrm{Tr}_{\mathcal{F}_{\Lambda }^{\prime }}\left\{ e^{-\beta \hat{H}%
_{\Lambda ,\lambda _{0}}^{SB}\left( c,\mu \right) }\right\} \leq \frac{1}{%
\beta V}\ln \mathrm{Tr}_{\mathcal{F}_{\Lambda }^{\prime }}\left\{ e^{-\beta
H_{\Lambda ,\lambda _{0}}^{SB}\left( c,\mu \right) }\right\} -\frac{1}{V}%
\left\langle \Delta \right\rangle _{\hat{H}_{\Lambda ,\lambda
_{0}}^{SB}(c,\mu )},  \label{pressure avec c upper bound-1}
\end{equation}
where
\begin{equation}
\left\langle -\right\rangle _{\hat{H}_{\Lambda ,\lambda _{0}}^{SB}\left(
c,\mu \right) }:=\frac{\mathrm{Tr}_{\mathcal{F}_{\Lambda }^{\prime }}\left\{
-e^{-\beta \hat{H}_{\Lambda ,\lambda _{0}}^{SB}\left( c,\mu \right)
}\right\} }{\mathrm{Tr}_{\mathcal{F}_{\Lambda }^{\prime }}\left\{ e^{-\beta
\hat{H}_{\Lambda ,\lambda _{0}}^{SB}\left( c,\mu \right) }\right\} }.
\end{equation}
In particular, since for $k\in \Bbb{R}^{3},$ $0\leq \lambda _{k}\leq \lambda
_{0}$ by our assumption (B) on the interaction potential, we obtain from the
inequality (\ref{pressure avec c upper bound-1}) together with (\ref
{pressure avec c upper bound-2}) that
\begin{eqnarray}
\frac{1}{\beta V}\ln \mathrm{Tr}_{\mathcal{F}_{\Lambda }^{\prime }}\left\{
e^{-\beta \hat{H}_{\Lambda ,\lambda _{0}}^{SB}\left( c,\mu \right) }\right\}
&\leq &\frac{1}{\beta V}\ln \mathrm{Tr}_{\mathcal{F}_{\Lambda }^{\prime
}}\left\{ e^{-\beta H_{\Lambda ,\lambda _{0}}^{SB}\left( c,\mu \right)
}\right\} +\frac{2|c|^{2}\lambda _{0}-\mu }{V}-\frac{\lambda _{0}}{V^{2}}
\nonumber \\
&&+\frac{2\lambda _{0}}{V^{2}}\stackunder{k\in \Lambda ^{*}\backslash
\left\{ 0\right\} }{\sum }\left\langle a_{k}^{*}a_{k}\right\rangle _{\hat{H}%
_{\Lambda ,\lambda _{0}}^{SB}\left( c,\mu \right) }.
\label{pressure avec c upper bound}
\end{eqnarray}
The last term can be explicitly computed. We omit the details. In fact, for
any $\mu \in \Bbb{R}$ one can check that
\begin{equation}
\frac{1}{V}\stackunder{k\in \Lambda ^{*}\backslash \left\{ 0\right\} }{\sum }%
\left\langle a_{k}^{*}a_{k}\right\rangle _{\hat{H}_{\Lambda ,\lambda
_{0}}^{SB}\left( c,\mu \right) }=\mathcal{O}\left( 1\right) \mathrm{\ as\;}%
\Lambda \uparrow \Bbb{R}^{3}.
\end{equation}
Therefore, from (\ref{pressure avec c upper bound}) together with (\ref
{limit sup 1}) and (\ref{important inequality}) one deduces that
\begin{equation}
p^{SB}(\beta ,\mu ,c)\leq \stackunder{\alpha \leq 0}{\inf }\left\{
p_{0}^{B}(\beta ,\alpha ,|c|^{2})+\frac{(\mu -\alpha )^{2}}{2\lambda _{0}}%
\right\} .
\end{equation}
Together with the lower bound (\ref{pressure avec c lower bound}), this
inequality proves the lemma. $\square $

\begin{lemma}[The free-energy density $f_{\Lambda }^{SB}(\beta ,y,c)$ in the thermodynamic limit]
\label{LDP-lemma1bis}\mbox{} \newline
For any $c\in \Bbb{C}$, $y\geq 0$ and $\beta >0$, the thermodynamic limit $%
f^{SB}(\beta ,y,c)$ of the free-energy density $f_{\Lambda }^{SB}(\beta ,y,c)
$ (\ref{definition of the partial free energy}) equals
\[
f^{SB}(\beta ,y,c):=\stackunder{\Lambda }{\lim }f_{\Lambda }^{SB}(\beta
,y,c)=f_{0}^{B}\left( \beta ,y,x\right) +\frac{\lambda _{0}}{2}\left(
y+x\right) ^{2},
\]
with $x=|c|^{2}\geq 0,$ and $f_{0}^{B}\left( \beta ,y,x\right) $ defined as
the Legendre-Fenchel transform of $p_{0}^{B}\left( \beta ,\alpha ,x\right) $
(\ref{eq 4.26}), cf. Theorem \ref{theorem LDP-0}.
\end{lemma}

\noindent \textit{Proof.} The pressure $p_{\Lambda }^{SB}(\beta ,\mu ,c)$ (%
\ref{definition of the pressure as function of c}) can be rewritten as
\begin{equation}
p_{\Lambda }^{SB}(\beta ,\mu ,c)=\frac{1}{\beta V}\ln \stackunder{\Bbb{R}_{+}%
}{\int }e^{\beta V\left( \mu y-f_{\Lambda }^{SB}\left( \beta ,y,c\right)
\right) }\nu _{\Lambda }\left( \mathrm{d}y\right) +\mu |c|^{2},
\end{equation}
with $\nu _{\Lambda }(\mathrm{d}y)$ defined in (\ref{definition of nu}). It
is then straightforward to check that the thermodynamic limit $p^{SB}(\beta
,\mu ,c)$ of $p_{\Lambda }^{SB}(\beta ,\mu ,c)$ (\ref{definition of the
pressure as function of c}) equals
\begin{equation}
p^{SB}(\beta ,\mu ,c)=\stackunder{y\geq 0}{\sup }\left\{ \mu y-f^{SB}\left(
\beta ,y,c\right) \right\} +\mu |c|^{2},
\end{equation}
with $f^{SB}(\beta ,y,c)<\infty $ for $y\geq 0$. The derivative of the
pressure $p^{SB}(\beta ,\mu ,c)$ is continuous as a function of $\mu $, cf.
Lemma \ref{LDP-lemma1} and (\ref{eq 4.26bis}). Thus, by using the {\textit{Tauberien
theorem}} proven in \cite{MinlosPovzner}, the existence of $p^{SB}(\beta ,\mu
,c)$ already implies the convexity of $f^{SB}(\beta ,y,c)$ as a function of $%
y\geq 0$. In particular, it yields that
\begin{equation}
f^{SB}\left( \beta ,y,c\right) =\stackunder{\mu \in \Bbb{R}}{\sup }\left\{
\mu (y+|c|^{2})-p^{SB}(\beta ,\mu ,c)\right\} \mathrm{\ for}\;y\geq 0.
\label{legendre transformation}
\end{equation}
By using the explicit form of $p^{SB}(\beta ,\mu ,c)$ given by Lemma \ref
{LDP-lemma1}, a straightforward computation then gives
\begin{equation}
f^{SB}\left( \beta ,y,c\right) =\stackunder{\alpha \leq 0}{\sup }\left\{
\alpha (y+|c|^{2})-p_{0}^{B}(\beta ,\alpha ,|c|^{2})\right\} +\frac{\lambda
_{0}}{2}\left( y+x\right) ^{2}.
\end{equation}
$\square $

\subsection{Large deviations principles\label{Section LDP}}

Let $\mathcal{X}$ denote a topological vector space. A lower semi-continuous
function $\mathrm{I}:\mathcal{X}\to [0,\infty ]$ is called a rate function%
\textit{\ }if $\mathrm{I}$ is not identical $\infty $ and has compact level
sets, i.e., if $\mathrm{I}^{-1}([0,m])=\{x\in \mathcal{X}:\mathrm{I}(x)\le
m\}$ is compact for any $m\ge 0$. A sequence $\left\{ X_{l}\right\}
_{l=1}^{+\infty }$ of $\mathcal{X}$-valued random variables $X_{l}$ or the
corresponding sequence $\left\{ \Bbb{P}_{l}\right\} _{l=1}^{+\infty }$ of
probability measures on the Borel subsets of $\mathcal{X}$ satisfy the large
deviations upper bound with speed\textit{\ }$a_{l}$ and rate function $%
\mathrm{I}$ if, for any closed subset $\mathcal{C}$ of $\mathcal{X}$,
\begin{equation}
\limsup_{l\to +\infty }\frac{1}{a_{l}}\ln \Bbb{P}_{l}\left( X_{l}\in
\mathcal{C}\right) =\limsup_{l\to \infty }\frac{1}{a_{l}}\ln \Bbb{P}%
_{l}\left( \mathcal{C}\right) \leq -\inf_{\mathcal{C}}\mathrm{I}\left(
x\right) ,  \label{LDPupper}
\end{equation}
and they satisfy the large deviations lower bound if, for any open subset $%
\mathcal{G}$ of $\mathcal{X}$,
\begin{equation}
\liminf_{l\to +\infty }\frac{1}{a_{l}}\ln \Bbb{P}_{l}\left( X_{l}\in
\mathcal{G}\right) =\limsup_{l\to \infty }\frac{1}{a_{l}}\ln \Bbb{P}%
_{l}\left( \mathcal{G}\right) \leq -\inf_{\mathcal{G}}\mathrm{I}\left(
x\right) .  \label{LDPlower}
\end{equation}
If both, upper and lower bound, are satisfied, one says that $\left\{
X_{l}\right\} _{l=1}^{+\infty }$ or $\left\{ \Bbb{P}_{l}\right\}
_{l=1}^{+\infty }$ satisfy a \textit{large deviations principle}. The
principle is called \textit{weak} if the upper bound in (\ref{LDPupper})
holds only for compact\textit{\ }sets $\mathcal{C}$. This notion easily
extends to the situation where the distribution of $X_{l}$ is not
normalized, but a sub-probability distribution only. Observe also that one
of the most important conclusions from a large deviations principle is
Varadhan's Lemma, which says that, for any bounded and continuous function $%
\varphi :\mathcal{X}\to \Bbb{R}$,
\[
\lim_{l\to +\infty }\frac{1}{a_{l}}\ln \int \exp \left( a_{l}\varphi \left(
X_{l}\right) \right) \mathrm{d}\Bbb{P}=-\inf_{x\in \mathcal{X}}\left\{
\mathrm{I}\left( x\right) -\varphi \left( x\right) \right\} .
\]
For a comprehensive treatment of the theory of large deviations, see \cite
{DemboZeitouni}.

\noindent \textbf{Acknowledgements}

\smallskip

\noindent We are most grateful to Joel Lebowitz for interest to this project, for his help and
inspiring discussions. We also wish to thank him for hospitality extended to us in the Rutgers
University (NJ) that allowed to discuss different
aspects of the paper. The present version of the manuscript was finalized during the
J.-B.Bru visit of the Centre de Physique Th\'{e}orique - Luminy
UMR-6207. He is thankful to Pierre Duclos for invitation to CPT.


\begin{thebibliography}{99}

\bibitem{LewisPuleZagrebnov1}  J.T. Lewis, J.V. Pul\'{e} and V.A. Zagrebnov,
The Large Deviation Principle for the Kac Distribution, {\textit{Helv. Phys.
Acta}} \textbf{61}: 1063-1078 (1988).

\bibitem{LebLenSpo}  J.L. Lebowitz, M. Lenci and H. Spohn, Large Deviations
for Ideal Quantum Systems, \textit{J. Math.Phys.} \textbf{41}: 1224-1243
(2000).

\bibitem{GallLebMast}  G. Gallavotti, J.L. Lebowitz and V. Mastropietro,
Large deviations in rarefied quantum gases, \textit{J. Stat. Phys.} \textbf{%
108}(5/6): 831-861 (2002).

\bibitem{AngelescuVerbeureZagrebnov1}  N. Angelescu, A. Verbeure and V.A.
Zagrebnov, On Bogoliubov's model of superfluidity, \textit{J. Phys. A:
Math.Gen.} \textbf{25}: 3473-3491 (1992).

\bibitem{ZagBru}  V.A. Zagrebnov and J.-B. Bru, The Bogoliubov Model of
Weakly Imperfect Bose Gas, \textit{Phys. Rep.} \textbf{350}: 291-434 (2001).

\bibitem{AngelescuVerbeureZagrebnov2}  N. Angelescu, A. Verbeure and V.A.
Zagrebnov, Superfluidity III, \textit{J. Phys. A: Math.Gen.} \textbf{30}:
4895-4913 (1997).

\bibitem{Adams-Bru1}  S. Adams and J.-B. Bru, Critical Analysis of the
Bogoliubov Theory of Superfluidity, \textit{Physica A} \textbf{332}: 60-78
(2004).

\bibitem{Adams-Bru1bis}  S. Adams and J.-B. Bru, Exact solution of the
AVZ-Hamiltonian in the grand canonical ensemble, \textit{Annales Henri Poincar%
\'{e}} \textbf{5}: 405-434 (2004).

\bibitem{Adams-Bru2}  S. Adams and J.-B. Bru, A New Microscopic Theory of
Superfluidity at all Temperatures, \textit{Annales Henri Poincar\'{e}}
\textbf{5}: 435-476 (2004).

\bibitem{Bru2006}  J.-B. Bru, Beyond the dilute Bose gas, \textit{Physica A}
\textbf{359}: 306-344 (2006).

\bibitem{BrattelliRobinson}  O. Brattelli and D.W. Robinson, \textit{%
Operator Algebras and Quantum Statistical Mechanics, Vol II}, 2nd ed.
Springer-Verlag, New York (1996).

\bibitem{Ruelle}  D. Ruelle, \textit{Statistical Mechanics: Rigorous Results}%
, Benjamin-Reading, New-York (1969).

\bibitem{BruZagrebnov6}  J.-B. Bru and V.A. Zagrebnov, On condensations in
the Bogoliubov Weakly Imperfect Bose-Gas, {\textit{J. Stat. Phys.}} \textbf{%
99}:1297 {(}2000).

\bibitem{BruZagrebnov8}  V.A. Zagrebnov and J.-B. Bru, The Bogoliubov Model
of Weakly Imperfect Bose Gas, \textit{Phys. Rep.} \textbf{350}:291 (2001).

\bibitem{Ginibre}  J. Ginibre, On the Asymptotic Exactness of the Bogoliubov
Approximation for many Bosons Systems, \textit{Commun. Math. Phys.} \textbf{8%
}: 26-51 (1968).

\bibitem{ReedSimon}  M. Reed and B. Simon,{\ \textit{Methods of Modern
Mathematical Physics}}\textit{, Vol. I: Functional Analysis,} Academic
Press, New York-London (1972).

\bibitem{Simon1}  B. Simon,{\ }\textit{The Statistical Mechanics of Lattice
Gases, Vol. I}, Princeton Univ. Press, Princeton (1993).

\bibitem{LiebSeiringerYngvason3}  E.H. Lieb, R. Seiringer{, and} J.
Yngvason, Justification of c-Number Substitutions in Bosonic Hamiltonians,
\textit{Phys. Rev. Lett.} \textbf{94}: 080401-1-4 (2005) .

\bibitem{DemboZeitouni}  A. Dembo and O. Zeitouni, \textit{Large Deviations
Techniques and Applications, }2nd ed. Springer-Verlag, New-York (1998).

\bibitem{MinlosPovzner}  {\ R.A. Minlos and A.Ja. Povzner}, {Thermodynamic
limit for entropy}, \textit{Trans. Moscow Math. Soc.} \textbf{17: }269
(1967).
\end{thebibliography}
\end{document}